\documentclass[10pt,conference]{IEEEtran}

\usepackage{epsfig}
\usepackage{graphicx}
\usepackage{cite}
\usepackage{amsmath}
\usepackage{amssymb}

\newcommand{\comment}[1]  {}
\def\BE{\begin{equation}}
\def\EE{\end{equation}}
\def\BEA{\begin{eqnarray}}
\def\EEA{\end{eqnarray}}

\newcommand\ie{{\textsl{i.e.\,}}}
\newcommand\eg{{\textsl{e.g.\,}}}
\newcommand\etal{{\textsl{et al.\, }}}

\newcommand\vb{{\bf b}}

\newcommand\vn{{\bf n}}

\newcommand\vx{{\bf x}}
\newcommand\vy{{\bf y}}
\newcommand\vz{{\bf z}}
\newcommand\mA{{\bf A}} 

\newcommand\mI{{\bf I}}

\newcommand\mR{{\bf R}}
\newcommand\mS{{\bf S}}

\begin{document}
\renewcommand{\baselinestretch}{0.97}
\title{Gaussian Belief Propagation Based\\ Multiuser Detection}

\author{
\authorblockN{Danny Bickson$^1$ and Danny Dolev}
\authorblockA{School of Computer Science and Engineering \\
Hebrew University of Jerusalem \\
Jerusalem 91904, Israel \\
\{daniel51,dolev\}@cs.huji.ac.il} \and
\authorblockN{Ori Shental$^1$, Paul H. Siegel and Jack K. Wolf}
\authorblockA{Center for Magnetic Recording Research \\
University of California - San Diego \\
La Jolla, CA 92093 \\
\{oshental,psiegel,jwolf\}@ucsd.edu}
}
%

\maketitle

\begin{abstract}\footnotetext[1]{Contributed equally to this work.\\Supported in part by NSF
Grant No.~CCR-0514859 and EVERGROW, IP 1935 of the EU Sixth
Framework.} In this work, we present a novel construction for
solving the linear multiuser detection problem using the Gaussian
Belief Propagation algorithm. Our algorithm yields an efficient,
iterative and distributed implementation of the MMSE detector.
Compared to our previous formulation, the new algorithm offers a
reduction in memory requirements, the number of computational
steps, and the number of messages passed. We prove that a
detection method recently proposed by Montanari \etal is an
instance of ours, and we provide new convergence results
applicable to both.
\end{abstract}

\section{Introduction}
Belief propagation (BP), also known as the sum-product algorithm,
is a powerful and efficient tool in solving, exactly or
approximately, inference problems in probabilistic graphical
models. The underlying essence of estimation theory is to detect a
hidden input to a channel from its observed output. The channel
can be represented as a certain graphical model, while the
detection of the channel input is equivalent to performing
inference in the corresponding graph.

The use of BP~\cite{BibDB:BookPearl} for detection purposes has
been proven to be very beneficial in several applications in
communications. For randomly-spread code-division multiple-access
(CDMA) in the large-system limit, Kabashima has introduced a
tractable BP-based multiuser detection (MUD) scheme, which
exhibits near-optimal error performance for binary-input additive
white Gaussian noise (BI-AWGN) channels~\cite{BibDB:Kabashima}.
This message-passing scheme has recently been extended to the case
where the ambient noise level is
unknown~\cite{BibDB:NS,BibDB:NS2}. As for sub-optimal detection,
the nonlinear soft parallel interference cancellation (PIC)
detector was reformulated by Tanaka and Okada as an approximate BP
solution~\cite{BibDB:TanakaOkada} to the MUD problem.

In contrast to the dense, fully-connected nature of the graphical
model of the non-orthogonal CDMA channel, a one-dimensional (1-D)
intersymbol interference (ISI) channel can be interpreted as a
cycle-free tree graph~\cite{BibDB:FactorGraph}. Thus, detection in
1-D ISI channels (termed equalization) can be performed in an
optimal maximum a-posteriori (MAP) manner via BP, also known in
this context as the forward/backward, or BCJR,
algorithm~\cite{BibDB:BCJR}. Also, Kurkoski
\etal~\cite{BibDB:Kurkoski}, \cite{BibDB:KurkoskiCorrection} have
proposed an iterative BP-like detection algorithm for 1-D ISI
channels that uses a parallel message-passing schedule and
achieves near-optimal performance.

For the intermediate regime of non-dense graphs but with many
relatively short loops, extensions of BP to two-dimensional ISI
channels have been considered by Marrow and
Wolf~\cite{BibDB:ConfWolf}, and recently Shental
\etal~\cite{BibDB:ShentalITW,BibDB:ShentalISIT,BibDB:ShentalIT}
have demonstrated the near-optimality of a generalized version of
BP for such channels. Recently, BP has been proved to
asymptotically achieve optimal MAP detection for sparse linear
systems with Gaussian
noise~\cite{BibDB:MontanariTse,BibDB:ConfWangGuo}, for example, in
CDMA with sparse spreading codes.

An important class of practical sub-optimal detectors is based on
linear detection. This class includes, for instance, the
conventional single-user matched filter (MF), decorrelator (a.k.a.
zero-forcing equalizer), linear minimum mean-square error (MMSE)
detector and many other detectors with widespread
applicability~\cite{BibDB:BookVerdu,BibDB:BookProakis}. In
general, linear detection can be viewed as the solution to a
(deterministic) set of linear equations describing the original
(probabilistic) estimation problem. Note that the mathematical
operation behind linear detection extends to other tasks in
communication, \eg, channel precoding at the
transmitter~\cite{BibDB:Precoding}.

Recently, linear detection has been explicitly linked to
BP~\cite{Allerton}, using a Gaussian belief propagation (GaBP)
algorithm. This allows for a distributed implementation of the
linear detector~\cite{BibDB:BicksonShental_ISIT1}, circumventing
the need of, potentially cumbersome, direct matrix inversion (via,
\eg, Gaussian elimination). The derived iterative framework was
compared quantitatively with `classical' iterative methods for
solving systems of linear equations, such as those investigated in
the context of linear implementation of 
CDMA
demodulation~\cite{grant99iterative,BibDB:TanRasmussen,BibDB:YenerEtAl}.
GaBP is shown to yield faster convergence than these standard
methods. Another important work is the BP-based MUD,
recently derived and analyzed by Montanari \etal
~\cite{BibDB:MontanariEtAl} for Gaussian input symbols.

There are several drawbacks to the linear detection technique
of~\cite{Allerton}. First, the input matrix $\mR_{n \times n} =
\mS_{n \times k}^T\mS_{k \times n}$ (the chip correlation matrix)
needs to be computed prior to running the algorithm. This
computation requires $n^2k$ operations. In case where the matrix
$\mS$ is sparse~\cite{BibDB:ConfWangGuo}, the matrix $\mR$ might
not no longer be sparse. Second, GaBP uses $2n^2$ memory to store
the messages. For a large $n$ this could be prohibitive.

In this paper, we propose a new construction that addresses those
two drawbacks. In our improved construction, given a
non-rectangular CDMA matrix $\mS_{n \times k}$, we compute the
MMSE detector $x = (\mS^T\mS + \Psi)^{-1}\mS^Ty$ where $\Psi$ is
the AWGN diagonal covariance matrix. We utilize the GaBP algorithm
which is an efficient iterative distributed algorithm. The new
construction uses only $2nk$ memory for storing the messages. When
$k \ll n$ this represents significant saving relative to the
$2n^2$ in our previously proposed algorithm. Furthermore, we do
not explicitly compute $\mS^T\mS$, saving an extra $n^2k$
overhead.

We show that Montanari's algorithm~\cite{BibDB:MontanariEtAl} is
an instance of our method. By showing this, we are able to prove
new convergence results for Montanari's algorithm. Montanari
proves that his method converges on normalized random-spreading
CDMA sequences, assuming Gaussian signaling. Using binary
signaling, he conjectures convergence to the large system limit.
Here, we extend Montanari's result, to show that his algorithm
converges also for non-random CDMA sequences when binary signaling
is used, under weaker conditions. Another advantage of our work is
that we allow different noise levels per bit transmitted.

The paper is organized as follows. Section~\ref{sec_linear}
formulates the problem of linear detection and presents the
distributed GaBP-based linear detection scheme.
Section~\ref{sec_new_const} describes a novel construction for
efficiently computing the MMSE detector. The relation to a factor
graph construction is explored in Section~\ref{sec_factor}. New
convergence results for Montanari's work are presented in
Section~\ref{sec_conv}. We conclude in Section~\ref{sec_end}. In
the Appendix we further explore the relation to Montanari's work.

We shall use the following notations. The operator $\{\cdot\}^{T}$
stands for a vector or matrix transpose, the matrix $\mI_{N}$ is a
$N\times N$ identity matrix, while the symbols $\{\cdot\}_{i}$ and
$\{\cdot\}_{ij}$ denote entries of a vector and matrix,
respectively. $\mathrm{N}(i)$ is the set of graph node connected
to node $i$.

\section{Linear Detection via Belief Propagation}\label{sec_linear}

Consider a discrete-time channel with a real input vector
\mbox{$\vx=\{x_{1},\ldots,x_{K}\}^{T}$} governed by an arbitrary
prior distribution, $P_{\vx}$, and a corresponding real output
vector
\mbox{$\vy=\{y_{1},\ldots,y_{K}\}^{T}=f\{\vx^{T}\}\in\mathbb{R}^{K}$}.
Here, the function $f\{\cdot\}$ denotes the channel
transformation. By definition, linear detection compels the
decision rule to be\BE\label{eq_gld}
\hat{\vx}=\Delta\{\vx^{\ast}\}=\Delta\{\mA^{-1}\vb\}, \EE where
$\vb=\vy$ is the $K\times 1$ observation vector and the $K\times
K$ matrix $\mA$ is a positive-definite symmetric matrix
approximating the channel transformation. The vector $\vx^{\ast}$
is the solution (over $\mathbb{R}$) to $\mA\vx=\vb$. Estimation is
completed by adjusting the (inverse) matrix-vector product to the
input alphabet, dictated by $P_{\vx}$, accomplished by using a
proper clipping function $\Delta\{\cdot\}$ (\eg, for binary
signaling $\Delta\{\cdot\}$ is the sign function).

For example, linear channels, which appear extensively in many
applications in communication and data storage systems, are
characterized by the linear relation\[ \vy=f\{\vx\}=\mR\vx+\vn,
\] where $\vn$ is a $K\times 1$ additive noise vector and
\mbox{$\mR=\mS^{T}\mS$} is a positive-definite symmetric matrix,
often known as the correlation matrix. The $N\times K$ matrix
$\mS$ describes the physical channel medium while the vector $\vy$
corresponds to the output of a bank of filters matched to the
physical channel $\mS$.

Assuming linear channels with AWGN with variance $\sigma^{2}$ as
the ambient noise, the general linear detection
rule~(\ref{eq_gld}) can describe known linear detectors.
For example ~\cite{BibDB:BookVerdu,BibDB:BookProakis}:
\begin{itemize}
                 \item The conventional matched filter (MF) detector is obtained by taking \mbox{$\mA\triangleq\mI_{K}$} and $\vb=\vy$. This detector is optimal, in the MAP-sense, for the case of zero cross-correlations, \ie, $\mR=\mI_{K}$, as happens for orthogonal CDMA or when there is no ISI effect.
\item The decorrelator (zero forcing equalizer) is achieved by
substituting \mbox{$\mA\triangleq\mR$} and $\vb=\vy$. It is
optimal in the noiseless case.
                 \item The linear minimum mean-square error (MMSE) detector can also be described by using \mbox{$\mA=\mR+\sigma^{2}\mI_{K}$}. This detector is 
known to be optimal when the input distribution $P_{\vx}$ is
Gaussian.

               \end{itemize}

In general, linear detection is suboptimal because of its
deterministic underlying mechanism (\ie, solving a given set of
linear equations), in contrast to other estimation schemes, such
as MAP or maximum likelihood, that emerge from an optimization
criterion.

In~\cite{Allerton}, linear detection, in its general
form~(\ref{eq_gld}), was implemented using an efficient
message-passing algorithm. The linear detection problem was
shifted from an algebraic to a probabilistic domain. Instead of
solving a deterministic vector-matrix linear equation, an
inference problem is solved in a graphical model describing a
certain Gaussian distribution function. Given the overall channel
matrix $\mR$ and the observation vector $\vy$, one knows how to
write explicitly $p(\vx)$ and the corresponding graph
$\mathcal{G}$ with edge potentials (`compatibility functions')
$\psi_{ij}$ and self-potentials (`evidence') $\phi_{i}$. These
graph potentials are determined according to the following
pairwise factorization of the Gaussian distribution $p(\vx)$ \[
p(\vx)\propto\prod_{i=1}^{K}\phi_{i}(x_{i})\prod_{\{i,j\}}\psi_{ij}(x_{i},x_{j}),\]
resulting in \mbox{$\psi_{ij}(x_{i},x_{j})\triangleq
\exp(-x_{i}R_{ij}x_{j})$} and
\mbox{$\phi_{i}(x_{i})=\exp\big(b_{i}x_{i}-R_{ii}x_{i}^{2}/2\big)$}.
The set of edges $\{i,j\}$ corresponds to the set of all non-zero
entries of $\mA$ for which $i>j$. Hence, we would like to
calculate the marginal densities, which must also be Gaussian,
\mbox{$p(x_{i})\sim\mathcal{N}(\mu_{i}=\{\mR^{-1}\vy\}_{i},P_{i}^{-1}=\{\mR^{-1}\}_{ii})$},
where $\mu_{i}$ and $P_{i}$ are the marginal mean and inverse
variance (a.k.a. precision), respectively. It is shown that the
inferred mean $\mu$ is identical to the desired solution $x^{\ast}
= \mR^{-1}\vy$. Table I lists the GaBP algorithm update rules.

\begin{table*}[htb!]\label{tab_summary}
\vspace{-0.5cm} \normalsize \caption{Computing $\mA^{-1}\vb$ via
GaBP. Online matlab implementation is provided
in~\cite{MatlabGABP}.} \centerline{
\begin{tabular}{|c|c|l|}
  \hline
  \textbf{\#} & \textbf{Stage} & \textbf{Operation}\\
  \hline
  1. & \emph{Initialize} & Compute $P_{ii}=A_{ii}$ and $\mu_{ii}=b_{i}/A_{ii}$.\\
  && Set $P_{ki}=0$ and $\mu_{ki}=0$, $\forall k\neq i$.\\ \hline
  2. & \emph{Iterate} & Propagate $P_{ki}$ and $\mu_{ki}$, $\forall k\neq i \;
\mbox{\rm such that} \; A_{ki}\neq0$.\\& & Compute $P_{i\backslash
j}=P_{ii}+\sum_{{k}\in\mathbb{N}(i) \backslash j} P_{ki}$ and
$\mu_{i\backslash j} = P_{i\backslash
j}^{-1}(P_{ii}\mu_{ii}+\sum_{k \in \mathrm{N}(i) \backslash j}
P_{ki}\mu_{ki})$.\\
  && Compute $P_{ij} = -A_{ij}P_{i\backslash j}^{-1}A_{ji}$ and $\mu_{ij} =
-P_{ij}^{-1}A_{ij}\mu_{i\backslash j}$.\\\hline
  3. & \emph{Check} & If $P_{ij}$ and $\mu_{ij}$ did not converge, return to
    \#2. Else, continue to \#4.\\\hline
  4. & \emph{Infer} & $P_{i}=P_{ii}+\sum_{{k}\in\mathrm{N}(i)}
P_{ki}$ , $\mu_{i}=P_{i}^{-1}(P_{ii}\mu_{ii}+\sum_{k \in
\mathrm{N}(i)} P_{ki}\mu_{ki})$.\\
  \hline
  5. & \emph{Decide} & $\hat{x}_{i}=\Delta\{\mu_{i}\}$ \\\hline
\end{tabular}} 
\end{table*}

\section{Distributed Iterative Computation of the MMSE Detector}
\label{sec_new_const} In this section, we efficiently extend the
applicability of the proposed GaBP-based solver for systems with
symmetric matrices~\cite{Allerton} to systems with any square
(\ie, also nonsymmetric) or rectangular matrix. We first construct
a new symmetric data matrix $\tilde{\mR}$ based on an arbitrary
(non-rectangular) matrix $\mS\in\mathbb{R}^{k\times n}$ \BE
\label{newR} \tilde{\mR}\triangleq\left(
  \begin{array}{cc}
    \mI_{k} & \mS^T \\
    \mS & -\Psi \\
  \end{array}
\right)\in\mathbb{R}^{(k+n)\times(k+n)}. \EE Additionally, we
define a new vector of variables
$\tilde{\vx}\triangleq\{\hat{\vx}^{T},\vz^{T}\}^{T}\in\mathbb{R}^{(k+n)\times1}$,
where $\hat{\vx}\in\mathbb{R}^{k\times1}$ is the (to be shown)
solution vector and $\vz\in\mathbb{R}^{n\times1}$ is an auxiliary
hidden vector, and a new observation vector
$\tilde{\vy}\triangleq\{\mathbf{0}^{T},\vy^{T}\}^{T}\in\mathbb{R}^{(k+n)\times1}$.

Now, we would like to show that solving the symmetric linear
system $\tilde{\mR}\tilde{\vx}=\tilde{\vy}$ and taking the first
$k$ entries of the corresponding solution vector $\tilde{\vx}$ is
equivalent to solving the original (not necessarily symmetric)
system $\mR\vx=\vy$. Note that in the new construction the matrix
$\tilde{\mR}$ is sparse again, and has only $2nk$ off-diagonal
nonzero elements. When running the GaBP algorithm we have only
$2nk$ messages, instead of $n^2$ in the previous construction.

Writing explicitly the symmetric linear system's equations, we get
\[     \hat{\vx}+\mS^T\vz=\mathbf{0}, \ \ \ \   \mS\hat{\vx}-\Psi \vz=\vy.     \]
Thus, \[ \hat{\vx}=\Psi^{-1}\mS^{T}(\vy-\mS\hat{\vx}), \] and
extracting $\hat{\vx}$ we have \[
\hat{\vx}=(\mS^{T}\mS+\Psi)^{-1}\mS^{T}\vy. \] Note, that when the
noise level is zero, $\Psi=0_{m \times m}$, we get the
Moore-Penrose pseudoinverse solution\[
\hat{\vx}=(\mS^{T}\mS)^{-1}\mS^{T}\vy=\mS^{\dag}\vy. \]

\section{Relation to factor graph}
\label{sec_factor} In this section we give an alternate proof of
the correctness of our construction. Given the inverse covariance
matrix $\tilde{\mR}$ defined in (\ref{newR}), and the shift vector
$ \tilde{\vx}$ we can derive the matching self and edge potentials
\[ \psi_{ij}(x_{i},x_{j})\triangleq \exp(-x_{i}R_{ij}x_{j}) \]
\[ \phi_{i}(x_{i})\triangleq \exp(-1/2 x_{i}R_{ii}^2 x_{i} - x_i y_i) \]
which is a factorization of the Gaussian system distribution
\[ p(\vx) \propto \prod_i \phi_i(x_i) \prod_{i,j} \psi_{ij}(x_i,
x_j) = \]
\[ = \prod_{i \le k} \phi_i(x_i) \prod_{i > k} \phi_i(x_i)
\prod_{i,j} \psi_{ij}(x_i, x_j) = \]
\[ = \prod_{i \le k} \overbrace{\exp( -\frac{1}{2} x_i^2 )}^{\mbox{ prior on x}} \prod_{i > k} \exp(-\frac{1}{2}
\Psi_i x_i^2 - x_i y_i) \prod_{i,j}
\exp(-x_{i}\overbrace{S_{ij}}^{ R_{ij}} x_{j})
\]

Next, we show the relation of our construction to a factor graph.
We will use a factor graph with $k$ nodes to the left (the bits
transmitted) and $n$ nodes to the right (the signal received),
shown in Fig. $1$. Using the definition
$\tilde{\vx}\triangleq\{\hat{\vx}^{T},\vz^{T}\}^{T}\in\mathbb{R}^{(k+n)\times1}$
the vector $\hat{\vx}$ represents the $k$ input bits and the
vector $\vz$ represents the signal received. Now we can write the
system probability as:
\[ p(\tilde{\vx}) \propto \int_{\hat{\vx}} \mathcal{N}(\hat{\vx};0,I) \mathcal{N}(\vz;S\hat{\vx}, \Psi) d\hat{\vx} \]
It is known that the marginal distribution over $\vz$ is:
\[ = \mathcal{N}(\vz; 0, \mS^T\mS + \Psi) \]
This distribution is Gaussian, with the following parameters:
\[ E(\vz|\hat{\vx}) = (\mS^T\mS + \Psi)^{-1}\mS^T\vy \]
\[ Cov(\vz|\hat{\vx}) = (\mS^T\mS + \Psi)^{-1} \]

\begin{figure}
\begin{center}
  \includegraphics[width=200pt]{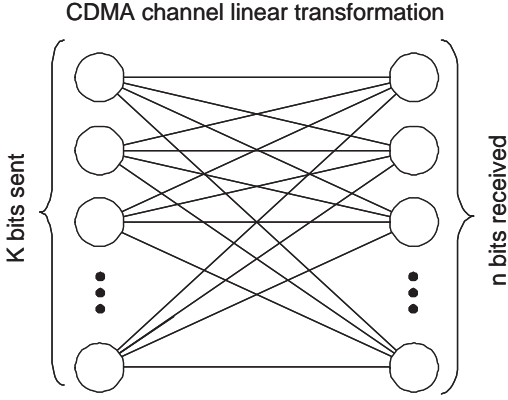}\\
  \caption{Factor graph describing the linear channel}\label{}
\end{center}
\end{figure}

 It is interesting to note that a similar
construction was used by Frey~\cite{frey99turbo} in his seminal
1999 work when discussing the factor analysis learning problem.
While his work is beyond the scope of this paper, it can be shown
that his algorithm can be modelled using the GaBP algorithm.

\section{New convergence results}
\label{sec_conv} One of the benefits of using our new construction
is that we propose a new mechanism to provide future convergence
results. In the Appendix we prove that Montanari's algorithm is an
instance of our algorithm, thus our convergence results apply to
Montanari's algorithm as well.

We know that if the matrix $\tilde{\mR}$ is strictly diagonally
dominant, then GaBP converges and the marginal means converge to
the true means~\cite[Claim 4]{BibDB:Weiss01Correctness}. Noting
that the matrix $\tilde{\mR}$ is symmetric, we can determine the
applicability of this condition by examining its columns.
Referring to (4) we see that in the first $k$ columns, we have the
$k$ CDMA sequences. We assume random-spreading binary CDMA
sequences which are normalized to one. In other words, the
absolute sum of each column is $\sqrt{n}$. In that case, the
matrix $\tilde{\mR}$ is not diagonally dominant (DD). We can add a
regularization term of $\sqrt{n} + \epsilon$ to force the matrix
$\tilde{\mR}$ to be DD, but we pay in changing the problem. In the
next $n$ columns of the matrix $\tilde{\mR}$, we have the diagonal
covariance matrix $\Psi$ with different noise levels per bit in
the main diagonal, and zero elsewhere. The absolute sum of each
column of $S$ is $k/\sqrt{n}$, thus when the noise level of each
bit satisfies $\Psi_i > k/\sqrt{n}$, we have a convergence
guarantee. Note, that the convergence condition is a {\em
sufficient condition}. Based on Montanari's work, we also know
that in the large system limit, the algorithm converges for binary
signaling, even in the absence of noise.

An area of future work is to utilize this observation to identify
CDMA schemes with matrices $\mS$ that when fitted into the matrix
$\tilde{\mR}$ are either DD, or comply to the spectral radius
convergence condition of~\cite{BibDB:jmw_walksum_nips}.
\section{Conclusion}
\label{sec_end} We presented a novel distributed algorithm for
computing the MMSE detector for the CDMA multiuser detection
problem. Our work utilizes the Gaussian Belief Propagation
algorithm while improving two existing
constructions~\cite{BibDB:MontanariEtAl,Allerton} in this field.
Although we described our algorithm in the context of multiuser
detection, it has wider applicability. For example, it provides an
efficient iterative method for computing the Moore-Penrose
pseudoinverse, and it can also be applied to the factor analysis
learning problem~\cite{frey99turbo}.

\section*{APPENDIX: Montanari's algorithm is an instance of our algorithm}
 In this section we show that Montanari's algorithm
is an instance of our algorithm. Our algorithm is more general.
First, we allow different noise level for each received bit,
unlike his work which uses a single fixed noise for the whole
system. In practice, the bits are transmitted using different
frequencies, thus suffering from different noise levels. Second,
the update rules in his paper are fitted only to the
randoml-spreading CDMA codes, where the matrix $A$ contains only
values which are drawn uniformly from $\{-1,1\}$. Assuming binary
signalling, he conjectures convergence to the large system limit.
Our new convergence proof holds for any CDMA matrices provided
that the absolute sum of the chip sequences is one, under weaker
conditions on the noise level. Third, we propose
in~\cite{Allerton} an efficient broadcast version for saving
messages in a broadcast supporting network.

The probability distribution of the factor graph used by Montanari
is:
\[ d\mu_y^{N,K} = \frac{1}{Z_y^{N,K}} \prod_{a=1}^{N}
\exp(-\frac{1}{2} \sigma^2 \omega_a^2 + j y_a \omega_a)
\prod_{i=1}^{K} \exp(-\frac{1}{2}x_i^2)  \]
\[ \cdot \prod_{i,a} \exp(
-\frac{j}{\sqrt{N}}s_{ai}\omega_a x_i) d\omega \]

Extracting the self and edge potentials from the above probability
distribution:
\[ \psi_{ii}(\vx_i) \triangleq \exp(-\frac{1}{2}\vx_i^2) \propto \mathcal{N}(\vx; 0, 1) \]
\[ \psi_{aa}(\omega_a) \triangleq \exp(-\frac{1}{2}\sigma^2 \omega^2_a +
j\vy_a \omega_a) \propto \mathcal{N}(\omega_a ; j \vy_a, \sigma^2) \]
\[ \psi_{ia}(\vx_i, \omega_a) \triangleq \exp(
-\frac{j}{\sqrt{N}}s_{ai}\omega_a \vx_i) \propto
\mathcal{N}(\vx; \frac{j}{\sqrt{N}}s_{ai}, 0) \]

For convenience, Table II provides a translation between the
notations used in this paper (taken from~\cite{Allerton}) and that
used by Montanari {\it et al.} in ~\cite{BibDB:MontanariEtAl}:
\begin{center}
\begin{table}[h!]
\caption{Summary of notations}
\begin{tabular}{|c|c|c|}
  \hline
  This work~\cite{Allerton} & Montanari {\it el al.}~\cite{BibDB:MontanariEtAl} & Description\\
  \hline
  $P_{ij}$ & $\lambda_{i \rightarrow a}^{(t+1)}$ & precision msg from left to right \\
  & $\hat{\lambda}_{a \rightarrow i}^{(t+1)}$ & precision msg from right to left \\
  $\mu_{ij}$ & $\gamma_{i \rightarrow a}^{(t+1)}$ & mean msg from left to right \\
  & $\hat{\gamma}_{a \rightarrow i}^{(t+1)}$ & mean msg from right to left  \\
  $\mu_{ii}$ & $y_i$ & prior mean of left node\\
  & $0$ & prior mean of right node\\
  $P_{ii}$ & $1$ & prior precision of left node\\
  $\Psi_{i}$ & $\sigma^2$ & prior precision of right node \\
  $\mu_i$ & $\frac{G_i}{L_i}$ & posterior mean of node \\
  $P_{i}$ & $L_i$ & posterior precision of node \\
  $A_{ij}$ & $\frac{-j s_{ia}}{\sqrt{N}}$ & covariance \\
  $A_{ji}$ & $\frac{-j s_{ai}}{\sqrt{N}}$ & covariance \\
      & $j$ & $j=\sqrt{-1}$ \\
  \hline
\end{tabular}
\end{table}
\end{center}

Now we derive Montanari's update rules. We start with the
precision message from left to right:
\[ \overbrace{\lambda_{i \rightarrow a}^{(t+1)}}^{P_{ij}} = 1 + \frac{1}{N} \Sigma_{b
\ne a} \frac{s_{ib}^2}{\lambda^{(t)}_{b \rightarrow i}}  = \]
\[ = \overbrace{1}^{P_{ii}} + \Sigma_{b \ne a} \overbrace{\frac{1}{N}
\frac{s_{ib}^2}{\lambda^{(t)}_{b \rightarrow i}}}^{P_{ki}} \]
\[ = \overbrace{1}^{P_{ii}} - \Sigma_{b \ne a} \overbrace{\frac{- j s_{ib}}{
\sqrt{N}}}^{-A_{ij}}
 \overbrace{\frac{1}{\lambda^{(t)}_{b \rightarrow i}}}^{(P_{j \backslash i})^{-1}} \overbrace{\frac{- j s_{ib}}{ \sqrt{N}}}^{A_{ji}} . \]
By looking at Table $1$, it is easy to verify that this precision
update rule is equivalent to that in \#2 of Table I.

Using the same logic we get the precision message from right to
left:
\[ \overbrace{\hat{\lambda}_{i \rightarrow a}^{(t+1)}}^{P_{ji}} = \overbrace{\sigma^2}^{P_{ii}} + \overbrace{\frac{1}{N}
\Sigma_{k \ne i} \frac{s_{ka}^2}{\lambda^{(t)}_{k \rightarrow
a}}}^{- A_{ij}^2 P_{j \backslash i}^{-1} }
\]

The mean message from left to right is given by
\[
\gamma_{i \rightarrow a}^{(t+1)} = \frac{1}{N} \Sigma_{b \ne a}
\frac{s_{ib}}{\lambda_{b \rightarrow i}^{(t)}}
\hat{\gamma}^{(t)}_{b \rightarrow i} = \]
\[ = \overbrace{0}^{\mu_{ii}}
-
 \Sigma_{b \ne a} \overbrace{\frac{- j s_{ib}}{
\sqrt{N}}}^{-A_{ij}} \overbrace{\frac{1}{\hat{\lambda}_{b
\rightarrow i}^{(t)}}}^{P_{j \backslash i}^{-1}}
\overbrace{\hat{\gamma}^{(t)}_{b \rightarrow i}}^{\mu_{j
\backslash i}} .
\]
The same calculation is done for the mean from right to left:
\[
\hat{\gamma}_{i \rightarrow a}^{(t+1)} = y_a - \frac{1}{N}
\Sigma_{k \ne i} \frac{s_{ka}}{\lambda_{k \rightarrow a}^{(t)}}
\gamma^{(t)}_{k \rightarrow a}
\]
Finally, the left nodes calculated the precision and mean by
\[
G_{i}^{(t+1)} = \frac{1}{\sqrt{N}} \Sigma_{b}
\frac{s_{ib}}{\lambda_{b \rightarrow i}^{(t)}}
\hat{\gamma}^{(t)}_{b \rightarrow i} \mbox{  ,  }  J_{i} =
G_{i}^{-1}
\]
\[ L_i^{(t+1)} =1 + \frac{1}{N} \Sigma_{b
} \frac{s_{ib}^2}{\lambda^{(t)}_{b \rightarrow i}} \mbox{   , }
 \mu_i = L_i G_i^{-1} . \]

The key difference between the two constructions is that Montanari
uses a directed factor graph while we use an undirected graphical
model. As a consequence, our construction provides additional
convergence results and simpler update rules.

\bibliographystyle{IEEEtran}   
\normalsize
\bibliography{IEEEabrv,CDMA,LDviaBP_Allerton07}       

\begin{thebibliography}{10}
\providecommand{\url}[1]{#1}
\csname url@rmstyle\endcsname
\providecommand{\newblock}{\relax}
\providecommand{\bibinfo}[2]{#2}
\providecommand\BIBentrySTDinterwordspacing{\spaceskip=0pt\relax}
\providecommand\BIBentryALTinterwordstretchfactor{4}
\providecommand\BIBentryALTinterwordspacing{\spaceskip=\fontdimen2\font plus
\BIBentryALTinterwordstretchfactor\fontdimen3\font minus
  \fontdimen4\font\relax}
\providecommand\BIBforeignlanguage[2]{{%
\expandafter\ifx\csname l@#1\endcsname\relax
\typeout{** WARNING: IEEEtran.bst: No hyphenation pattern has been}%
\typeout{** loaded for the language `#1'. Using the pattern for}%
\typeout{** the default language instead.}%
\else
\language=\csname l@#1\endcsname
\fi
#2}}

\bibitem{BibDB:BookPearl}
J.~Pearl, \emph{Probabilistic Reasoning in Intelligent Systems: Networks of
  Plausible Inference}.\hskip 1em plus 0.5em minus 0.4em\relax San Francisco:
  Morgan Kaufmann, 1988.

\bibitem{BibDB:Kabashima}
Y.~Kabashima, ``A {CDMA} multiuser detection algorithm on the basis of belief
  propagation,'' \emph{J. Phys. A: Math. Gen.}, vol.~36, pp. 11\,111--11\,121,
  Oct. 2003.

\bibitem{BibDB:NS}
J.~P. Neirotti and D.~Saad, ``Improved message passing for inference in densely
  connected systems,'' \emph{Europhys.~Lett.}, vol.~71, no.~5, pp. 866--872,
  Sept. 2005.

\bibitem{BibDB:NS2}
------, ``Efficient {Bayesian} inference for learning in the {Ising} linear
  perceptron and signal detection in {CDMA},'' \emph{Physica A}, vol. 365, pp.
  203--210, Feb. 2006.

\bibitem{BibDB:TanakaOkada}
T.~Tanaka and M.~Okada, ``Approximate belief propagation, density evolution,
  and statistical neurodynamics for {CDMA} multiuser detection,'' \emph{{IEEE}
  Trans. Inform. Theory}, vol.~51, no.~2, pp. 700--706, Feb. 2005.

\bibitem{BibDB:FactorGraph}
F.~Kschischang, B.~Frey, and H.~A. Loeliger, ``Factor graphs and the
  sum-product algorithm,'' \emph{{IEEE} Trans. Inform. Theory}, vol.~47, pp.
  498--519, Feb. 2001.

\bibitem{BibDB:BCJR}
L.~R. Bahl, J.~Cocke, F.~Jelinek, and J.~Raviv, ``Optimal decoding of linear
  codes for minimizing symbol error rate,'' \emph{{IEEE} Trans. Inform.
  Theory}, vol.~20, no.~3, pp. 284--287, Mar. 1974.

\bibitem{BibDB:Kurkoski}
B.~M. Kurkoski, P.~H. Siegel, and J.~K. Wolf, ``Joint message-passing decoding
  of {LDPC} codes and partial-response channels,'' \emph{{IEEE} Trans. Inform.
  Theory}, vol.~48, pp. 1410--1422, June 2002.

\bibitem{BibDB:KurkoskiCorrection}
------, ``Correction to '{J}oint message-passing decoding of {LDPC} codes and
  partial-response channels','' \emph{{IEEE} Trans. Inform. Theory}, vol.~49,
  p. 2076, Aug. 2003.

\bibitem{BibDB:ConfWolf}
M.~Marrow and J.~K. Wolf, ``Iterative detection of $2$-dimensional {ISI}
  channels,'' in \emph{Proc. {IEEE} Inform. Theory Workshop ({ITW})}, Paris,
  France, Mar. 2003, pp. 131--134.

\bibitem{BibDB:ShentalITW}
O.~Shental, N.~Shental, A.~J. Weiss, and Y.~Weiss, ``Generalized belief
  propagation receiver for near-optimal detection of two-dimensional channels
  with memory,'' in \emph{Proc. {IEEE} Information Theory Workshop ({ITW})},
  San Antonio, Texas, USA, Oct. 2004, pp. 225--229.

\bibitem{BibDB:ShentalISIT}
O.~Shental, N.~Shental, and \mbox{S. Shamai (Shitz)}, ``On the achievable
  information rates of two-dimensional channels with memory,'' in \emph{Proc.
  {IEEE} Int. Symp. Inform. Theory ({ISIT})}, Adelaide, Australia, Sept. 2005.

\bibitem{BibDB:ShentalIT}
O.~Shental, N.~Shental, \mbox{S. Shamai (Shitz)}, I.~Kanter, A.~J. Weiss, and
  Y.~Weiss, ``Finite-state input two-dimensional {Gaussian} channels with
  memory: Estimation and information via graphical models and statistical
  mechanics,'' \emph{{IEEE} Trans. Inform. Theory}, vol.~54, pp. 1500 -- 1513,
  April 2008.

\bibitem{BibDB:MontanariTse}
A.~Montanari and D.~Tse, ``Analysis of belief propagation for non-linear
  problems: The example of {CDMA} (or: How to prove {Tanaka's} formula),'' in
  \emph{Proc. {IEEE} Inform. Theory Workshop ({ITW})}, {Punta del Este},
  Uruguay, Mar. 2006.

\bibitem{BibDB:ConfWangGuo}
C.~C. Wang and D.~Guo, ``Belief propagation is asymptotically equivalent to
  {MAP} detection for sparse linear systems,'' in \emph{Proc. 44th Allerton
  Conf. on Communications, Control and Computing}, Monticello, IL, USA, Sept.
  2006.

\bibitem{BibDB:BookVerdu}
S.~Verd\'{u}, \emph{Multiuser Detection}.\hskip 1em plus 0.5em minus
  0.4em\relax Cambridge, UK: Cambridge University Press, 1998.

\bibitem{BibDB:BookProakis}
J.~G. Proakis, \emph{Digital Communications}, 4th~ed.\hskip 1em plus 0.5em
  minus 0.4em\relax New York, USA: McGraw-Hill, 2000.

\bibitem{BibDB:Precoding}
B.~R. Voj\v{c}i\'{c} and W.~M. Jang, ``Transmitter precoding in synchronous
  multiuser communications,'' \emph{{IEEE} Trans. Commun.}, vol.~46, no.~10,
  pp. 1346--1355, Oct. 1998.

\bibitem{Allerton}
D.~Bickson, O.~Shental, P.~H. Siegel, J.~K. Wolf, and D.~Dolev, ``Linear
  detection via belief propagation,'' in \emph{Proc. 45th Allerton Conf. on
  Communications, Control and Computing}, Monticello, IL, USA, Sept. 2007.

\bibitem{BibDB:BicksonShental_ISIT1}
O.~Shental, D.~Bickson, P.~H. Siegel, J.~K. Wolf, and D.~Dolev, ``Gaussian
  belief propagation solver for systems of linear equations,'' in \emph{{IEEE}
  Int. Symp. Inform. Theory ({ISIT})}, Toronto, Canada, July 2008.

\bibitem{grant99iterative}
A.~Grant and C.~Schlegel, ``Iterative implementations for linear multiuser
  detectors,'' \emph{{IEEE} Trans. Commun.}, vol.~49, no.~10, pp. 1824--1834,
  Oct. 2001.

\bibitem{BibDB:TanRasmussen}
P.~H. Tan and L.~K. Rasmussen, ``Linear interference cancellation in {CDMA}
  based on iterative techniques for linear equation systems,'' \emph{{IEEE}
  Trans. Commun.}, vol.~48, no.~12, pp. 2099--2108, Dec. 2000.

\bibitem{BibDB:YenerEtAl}
A.~Yener, R.~D. Yates, , and S.~Ulukus, ``{CDMA} multiuser detection: {A}
  nonlinear programming approach,'' \emph{{IEEE} Trans. Commun.}, vol.~50,
  no.~6, pp. 1016--1024, June 2002.

\bibitem{BibDB:MontanariEtAl}
A.~Montanari, B.~Prabhakar, and D.~Tse, ``Belief propagation based multi-user
  detection,'' in \emph{Proc. 43th Allerton Conf. on Communications, Control
  and Computing}, Monticello, IL, USA, Sept. 2005.

\bibitem{MatlabGABP}
{Gaussian Belief Propagation} implementation in matlab [online] {\tt
  http://www.cs.huji.ac.il/labs/danss/p2p/gabp/}.

\bibitem{frey99turbo}
B.~J. Frey, ``Local probability propagation for factor analysis,'' in
  \emph{Neural Information Processing Systems ({NIPS})}, Denver, Colorado, Dec
  1999, pp. 442--448.

\bibitem{BibDB:Weiss01Correctness}
Y.~Weiss and W.~T. Freeman, ``Correctness of belief propagation in {Gaussian}
  graphical models of arbitrary topology,'' \emph{Neural Computation}, vol.~13,
  no.~10, pp. 2173--2200, 2001.

\bibitem{BibDB:jmw_walksum_nips}
J.~K. Johnson, D.~M. Malioutov, and A.~S. Willsky, ``Walk-sum interpretation
  and analysis of {Gaussian} belief propagation,'' in \emph{Advances in Neural
  Information Processing Systems 18}, Y.~Weiss, B.~Sch\"{o}lkopf, and J.~Platt,
  Eds.\hskip 1em plus 0.5em minus 0.4em\relax Cambridge, MA: MIT Press, 2006,
  pp. 579--586.

\end{thebibliography}

\end{document}